# Photonuclear reactions in astrophysics

*T. Rauscher*
*Department of Physics, University of Basel, 4056 Basel, Switzerland*
*and*
*Centre for Astrophysics Research, University of Hertfordshire, Hatfield AL10 9AB, United Kingdom*

## Photodisintegration in stellar environments

Nucleosynthesis in stars and stellar explosions proceeds via nuclear reactions in thermalized plasmas. Nuclear reactions not only transmute elements and their isotopes, and thus create all known elements from primordial hydrogen and helium, they also release energy to keep stars in hydrostatic equilibrium over astronomical timescales. A stellar plasma has to be hot enough to provide sufficient kinetic energy to the plasma components to overcome Coulomb barriers and to allow interactions between them. Plasma components in thermal equilibrium are bare atomic nuclei, free electrons, and photons (radiation). Typical temperatures of plasmas experiencing nuclear burning range from $10^7$ K for hydrostatic hydrogen burning (mainly interactions among protons and He isotopes) to $10^{10}$ K or more in explosive events, such as supernovae or neutron star mergers. This still translates into low interaction energies by nuclear physics standards, as the most probable energy $E$ between reaction partners in terms of temperature is derived from Maxwell-Boltzmann statistics and yields $E = T_9/11.6045$ MeV, where $T_9$ is the plasma temperature in GK.

Photodisintegration reactions only significantly contribute when the plasma temperature is sufficiently high to have an appreciable number of photons (given by a Planck radiation distribution) at energies exceeding the energy required to separate neutrons, protons, and/or α particles from a nucleus. Forward and reverse reactions are always competing in a stellar plasma and thus photodisintegrations have to be at the same level or faster than capture reactions in order to affect nucleosynthesis. Since the number of captures per second and volume (the capture *rate*) not only scales with temperature but also with plasma density [1], the threshold temperature at which photodisintegrations cannot be neglected is higher for denser plasmas. On the other hand, photons require less energy when the particle separation energies are small. This is the case when approaching the neutron- or proton-dripline or for (γ,α) reactions in the region of spontaneous α-emitters. Based on the reciprocity relation for nuclear reactions, the principle of detailed balance can be derived, which relates the reactivity of the forward reaction (capture) $r_f^*$ to the reaction rate of the reverse reaction $\lambda_r^*$ (photodisintegration) [1]. Apart from factors containing spin weights and reduced masses, $\lambda_r^*$ is proportional to $T_9^{3/2} \exp(-11.6045 Q/T_9) r_f^*$, where $Q$ is the reaction Q-value given in MeV, describing the energy release of the forward reaction. This means that mainly the Q-value sets the temperature at which the reverse reaction becomes fast enough to compete with the forward reaction and affect the amount of a given nuclide in the stellar plasma. The Q-value of a capture reaction is just the separation energy of the projectile in the final nucleus.

Due to the above relations, photodisintegrations are found to be important in roughly three contexts:
1. (Almost) complete photodisintegration at the onset of a hydrostatic burning phase;
2. Partial photodisintegration in explosive burning;
3. Reaction equilibria between captures and photodisintegrations in explosive burning when both types of rates are fast and affecting nuclear abundances at a timescale much shorter than the process duration.

Regarding case 1), a star with more than 8 – 10 times the solar mass evolves through a series of hydrostatic burning phases named after the main element used up as "nuclear fuel" for energy generation. These phases are, in this order, H-, He-, C-, Ne-, O-, and Si-burning. There is no stable burning phase after Si-burning but the core of the star collapses to essentially nuclear density, turning into a neutron star. As a consequence of this collapse, thermal and kinetic energy is deposited into the outer layers of the star, causing a shockwave running outwards and ejecting most of these layers. This is called a core-collapse supernova. The hydrostatic burning phases leading up to the final collapse mainly involve fusion reactions, including fusion of the main fuel and additional captures of protons and/or α-particles on nuclides along stability up to the Fe-region. Free neutrons do not play a role in regular hydrostatic burning but are released in larger amounts in the He-shells of thermally pulsing AGB stars, where they cause the production of nuclei heavier than Fe by neutron captures along stability. This is called the main s-process [2]. The reaction Q-values of all these captures along stability are of the order of a few tens MeV and so the reverse reactions—the photodisintegrations—do not play a role in hydrostatic burning and s-processing because the plasma temperatures are not high enough. The only two exceptions are photodisintegration of $^{20}$Ne, which initiates the Ne-burning phase of a star, and photodisintegration of $^{28}$Si at the start of the Si-burning phase. The α-separation energy in $^{20}$Ne is only 4.73 MeV, permitting the destruction of $^{20}$Ne before other reactions set in. Typical conditions of hydrostatic Ne-burning are 1.2 GK at a density of 1000 kg/cm$^3$. The released α-particles then react with the remaining nuclides by (α,γ), (α,n), and (α,p) reactions leading the way to further reactions with the released nucleons, and so on. A similar destruction of $^{28}$Si occurs at the start of the Si-burning phase, when $^{28}$Si is destroyed by (γ,n), (γ,p), (γ,α) reactions at about 3 – 4 GK, giving rise to a suite of subsequent reactions. At such temperatures and Si-burning densities of about $10^4$ kg/cm$^3$, reaction equilibria as mentioned above in case 3) can already be established. This implies that after the initial destruction of $^{28}$Si the resulting abundances of nuclides in the plasma are not determined by individual reactions anymore but rather by equilibrium abundances. This is explained in more detail below.

Regarding case 2), the prerequisites for partial destruction of nuclei by photodisintegration are sufficiently high temperature but also lower densities than those in Si-burning and a short process duration to avoid complete destruction. Such conditions are found in explosive Ne/O-burning of a core-collapse supernova, when the supernova shockwave passes the Ne- and O-layers of the star, raising the temperature to 2 – 4 GK for a few seconds. Similar conditions are also found in simulations of the thermonuclear explosion of a white dwarf, called type Ia supernova or thermonuclear supernova (not to be confused with a core-collapse supernova). In such a thermonuclear supernova, some regions of the white dwarf, which is completely disrupted in the explosion, experience thermodynamic conditions suited for partial disintegration of the nuclei contained therein. These two sites have been suggested to be the source of the so-called *p-nuclides*, which are 32 proton-rich isotopes of elements from Sr to Hg [3]. Their existence cannot be explained by neutron captures in the s- and r-processes (producing all other nuclides beyond Fe) and thus require a different production mechanism. These proton-rich isotopes can be reached by sequences of (γ,n) reactions on stable nuclei pre-existing in the hot plasma. These (γ,n) reactions then compete with (γ,p) (below neutron number 82) and (γ,α) (for $N \geq 82$) when reaching unstable nuclei on the proton-rich side of stability.

Regarding case 3), a reaction equilibrium can be established when both forward and reverse reactions are sufficiently fast to alter the abundance of a nuclide during the nucleosynthesis period. Process timescales are of the order of a few seconds in explosive nucleosynthesis and thus the reaction rates have to allow for a significant number of reactions during that time in both reaction directions. If the equilibrium can be established, the abundances of the nuclides in the plasma assume their equilibrium value, which is independent of the actual rates but depends on temperature, density, nuclear binding energy or separation energy, and the neutron-to-proton ratio [1]. A full *nuclear statistical equilibrium* (NSE) with all reactions, except for those mediated by the weak interaction, being in equilibrium is encountered in the innermost parts of a star which are barely ejected in a core-collapse supernova. It can also be established in some parts of hot accretion disks around black holes and in neutron star mergers.

We can also distinguish several types of partial equilibria. One type is found in hydrostatic and explosive Si-burning, where groups of nuclei are in equilibrium and the groups are connected by slower reactions not in equilibrium (*quasi-statistical equilibrium*, QSE). An equilibrium between neutron captures and (γ,n) reactions is found in the *hot r-process* occurring in hot, neutron-rich matter ejected in neutron star mergers and possibly also in magnetized jets formed in some types of core-collapse supernova explosions [4,5]. Here, the isotope abundances in each isotope chain are set by the (n,γ)-(γ,n) equilibrium abundances and β$^-$-decays connect neighbouring isotope chains. As the equilibration within a chain of nuclei is much faster than any β$^-$-decay half-life even far off stability, the most abundant nuclei were often called "waiting points", as the r-process flow to heavier elements is halted until they decay. In hot, proton-rich environments, a (p,γ)-(γ,p) equilibrium can be found, involving proton-rich, unstable nuclides. Such environments include proton-rich, dense inner zones of a core-collapse supernova which are ejected under influence of a neutrino wind, giving rise to a *νp-*

*process* [6]. Another site developing a (p,γ)-(γ,p) equilibrium is thermonuclear burning on the surface of a neutron star, causing *X-ray bursts* [7]. Due to the Coulomb barriers, the most abundant nuclei in the isotonic chains are not as far off from stability as the ones in the r-process, although higher temperature is required to establish this equilibrium. Furthermore, the extension of the production chains to heavier elements is more limited than in the r-process, not only due to increasing Coulomb barriers but also because of the region of spontaneous α-emitters found on the proton-rich side of the nuclear chart at higher mass numbers. When the reaction flow runs into this region, α-emission will be faster than proton emission, limiting the further progress [8].

Common to all these equilibria is the fact that photodisintegration reactions compete with capture reactions but at the same time it is not necessary to know the individual capture and photodisintegration rates to determine the nuclear abundances. The rates cancel out in the final expression for the abundance of a nucleus due to the reciprocity relation between forward and reverse rate.

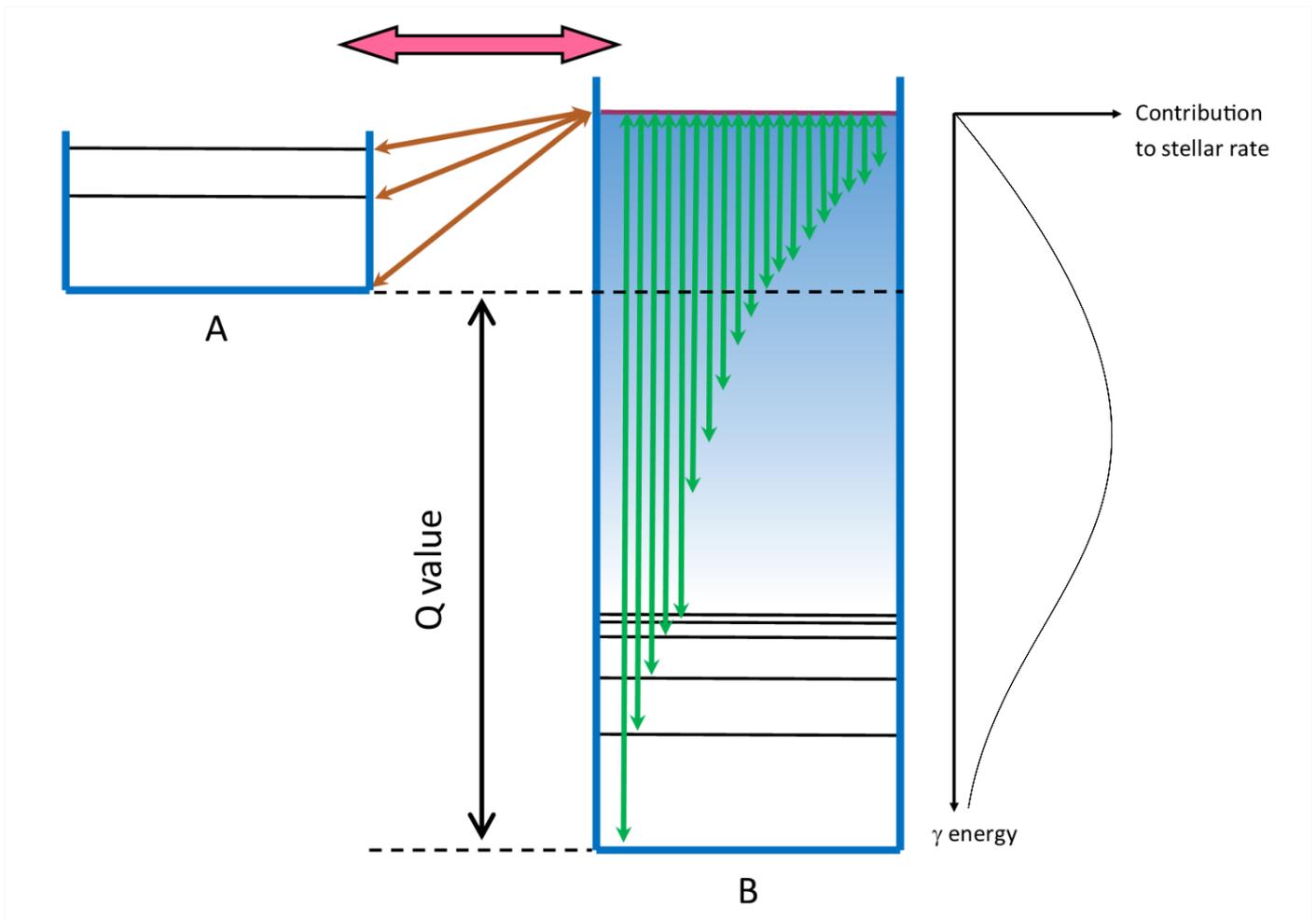

*Figure 1: Sketch of particle transitions (brown arrows) and γ-transitions (green arrows) contributing to the stellar rate of a reaction A+x ↔ B+γ, proceeding via a compound state (red). Under stellar conditions forward and reverse reactions are in equilibrium and reactions commence also from excited states in nucleus A and B.*

## Photodisintegration experiments for astrophysical applications

As outlined above, there is only a limited number of reactions for which the rates have to be known explicitly. These are the ones contributing to the production of p-nuclides, as outlined in case 2). It is tempting to assume that the required cross sections could be determined directly through a laboratory measurement. This is not feasible, however. The problem is not so much the fact that many of the relevant reactions involve unstable nuclei but rather that reactions on excited states of nuclei have to be included in the calculation of the *stellar* rates whereas experiments only measure cross sections of nuclei in the ground state [9]. The situation is sketched in **Figure 1**. In a reaction *A+x ↔ B+γ* occurring in a stellar plasma, the nuclear states in nuclides *A* and *B* are connected by particle- and γ-transitions in both reaction directions. Excited states are bombarded by particles and photons with energies given by a Maxwell-Boltzmann and Planck distribution, respectively, at plasma temperature *T*. and therefore also reactions on excited states contribute significantly to the total stellar rate. As can be inferred from **Figure 1**, the reaction in the direction of positive reaction Q-value is affected less by contributions from excited states. Moreover, it was shown that the contribution of excited states is larger by orders of magnitude for photodisintegrations than for captures, even for endothermic captures [1,10-12]. Furthermore, the higher the plasma temperature, the more excited states contribute. This worsens the situation since photodisintegrations are only important in nucleosynthesis at high temperature. The ground state contribution to the stellar rate for reactions concerning p-nuclides typically is only a few tenths of a per mille [13]. Thus, a determination of the photonuclear cross section only determines a small fraction of the actual stellar rate and is not suited to constrain the rate.

Photodisintegration experiments can only be used to derive information on certain nuclear properties required for the calculation of the stellar rates and thus to test and support theory [9]. One typical goal of an experiment is the determination of the E1 photon strength function determining the strength of the γ-transitions between excited states which enter the calculation of the rate. However, only photon energies larger than the particle separation energy can be probed when studying photo-induced particle emission. The really interesting energy range would, however, be at about half of that (typically 3 – 4 MeV), due to the necessity of folding the strength function with the number of available final states, as given by the nuclear level density, in the calculation of the total rate [14]. (The relative importance of γ-energies contributing to the stellar rate is also shown in **Figure 1**.) Nevertheless, the knowledge of the position of the peak of the E1 Giant Dipole Resonance (GDR), which can be determined at higher photon energy is a valuable additional piece of knowledge to normalize theoretically predicted GDR energies. A combination of (γ,γ') data with theoretical considerations has been used in the determination of stellar rates [15].

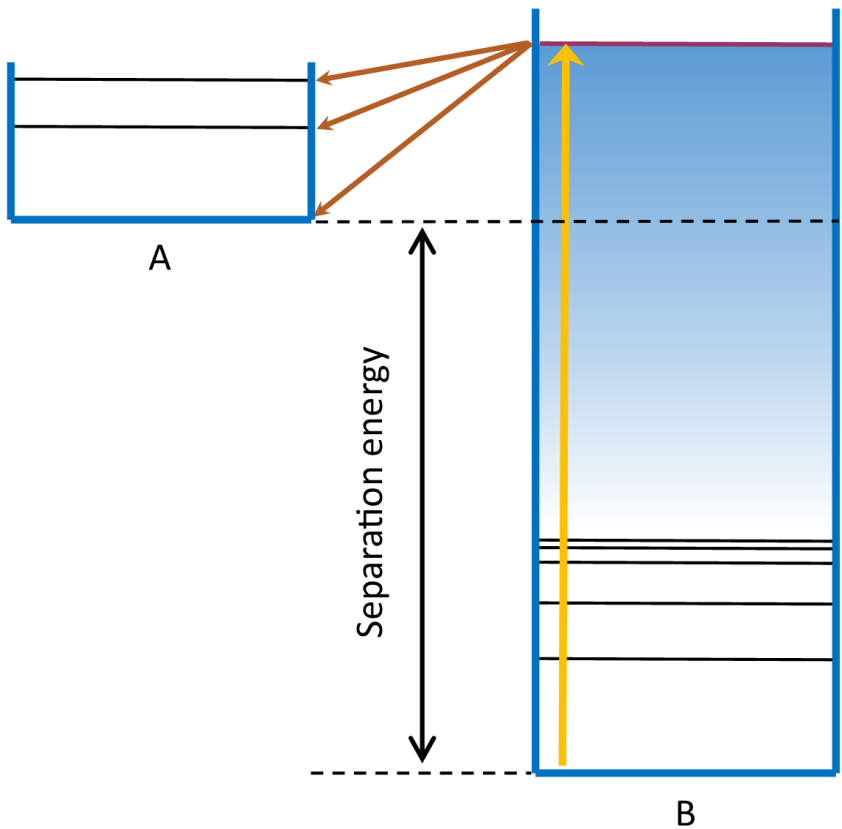

*Figure 2: Photodisintegration experiments can probe particle emission transitions to ground and excited states.*

An interesting application of (γ,n), (γ,p), and (γ,α) experiments, respectively, is the determination of the relative strengths of particle emissions to excited states of the final nucleus. Such an experiment is sketched in **Figure 2**. It serves the purpose of testing the predictions of transitions from and to excited target states in stellar capture reactions. One has to be careful in the interpretation of such experiments, however, because starting from a specific ground state introduces a selection of possible quantum numbers (relative angular momenta) which may be different from the astrophysically relevant ones. Nevertheless, in testing the predicted ratios relative to the ground state transitions, e.g., $(\gamma,n_1)/(\gamma,n_0)$, $(\gamma,n_2)/(\gamma,n_0)$, or $(\gamma,\alpha_1)/(\gamma,\alpha_0)$, $(\gamma,\alpha_2)/(\gamma,\alpha_0)$, and so on, the dependence on the E1 strength cancels out. Such a probing of particle transitions may help to improve the reaction models used to predict reaction rates [9].

The number of accessible states, and thus of contributing transitions, is reduced in nuclides with a low nuclear level density at the separation energy of particle $x$ in the compound nucleus $B$. In this case direct capture into a final state is more probable than a reaction via a compound state. The energy difference between the initial system $A+x$ and the final system $B$ is emitted as a single γ-ray. In analogy, the reverse reaction can also proceed directly by photo-induced emission of the particle $x$. In the stellar environment both reaction directions and all initial and final states accessible under consideration of quantum mechanical selection rules and available energy have to be considered, as sketched in **Figure 3**. Nevertheless, a photodisintegration experiment measuring a *direct reaction* starting with a nucleus in the ground state may subsume a larger fraction of the total stellar rate compared to the case of compound reactions discussed above. Direct reactions are important for the lightest nuclides (mass number A≤10) or for heavier nuclides with low particle separation energies, e.g., close to the proton- or neutron-dripline. Even when astrophysical photodisintegration is not of immediate importance for most of such nuclei (reaction equilibria are established when reaching conditions to synthesise nuclides close to the driplines), a photodisintegration experiment may still provide additional information on a considerable fraction of the contributing transitions and could be combined with theory to obtain an improved constraint of a stellar direct capture rate.

Despite of the restrictions with respect to astrophysical reaction rates, photonuclear experiments can complement other types of measurements, to provide nuclear structure information also for astrophysical application as well as further test cases for models predicting nuclear properties and cross sections. Not discussed here as it concerns completely different process conditions in astro-particle physics, the knowledge of photo-induced reaction cross sections at high energy is also essential for the investigation of the origin and the propagation of Galactic Cosmic Rays, involving nuclei moving at relativistic velocities through a radiation background (see, e.g., [16] and references therein).

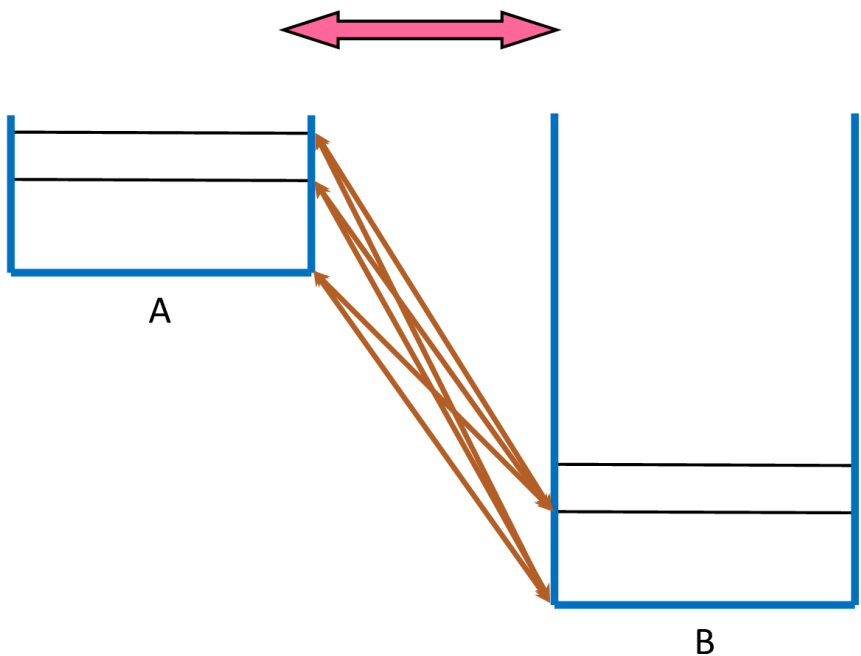

*Figure 3: In a direct reaction the projectile is captured directly into a final state under emission of the energy difference between initial and final configuration as a γ-quant. Direct reactions involve fewer transitions than compound reactions. This reaction mechanism dominates for nuclides with low nuclear level density at the particle separation energy, such as in the lightest nuclei or isotopes close to the proton- or neutron-dripline.*